\documentclass[a4paper]{article}
\usepackage{RR}
\RRNo{6242}
\usepackage{hyperref}
\RRdate{July 2007}
\RRauthor{Yves Bertot}
\authorhead{Bertot}
\RRetitle{%
Theorem proving support in programming language semantics
}
\RRtitle{%
S\'emantique des langages de programmation avec le support
d'un outil de preuve
}
\RRabstract{We describe several views of the semantics of a simple
programming language as formal documents in the calculus
of inductive constructions that can be verified by the Coq proof
system.  Covered aspects are natural semantics,
 denotational semantics, axiomatic semantics, and
abstract interpretation.
Descriptions as recursive functions are also provided whenever suitable,
thus yielding a a verification condition generator and a
static analyser that can be run inside the theorem prover for use
in reflective proofs.  Extraction of an interpreter from the
denotational semantics is also described.  All different aspects are
formally proved sound with respect to the natural semantics
specification.}
\RRresume{Nous d\'ecrivons plusieurs points de vue sur la s\'emantique
d'un petit langage de programmation, vus comme des documents dans
le calcul des constructions inductives
qui peuvent \^etre v\'erifi\'es par le
syst\`eme Coq.  Les aspects couverts sont la s\'emantique naturelle,
la s\'emantique d\'enotationnelle, la s\'emantique axiomatique, et
l'interpr\'etation abstraite.  Des descriptions sous formes de fonctions
r\'ecursives sont fournies quand c'est adapt\'e, et on obtient ainsi
un g\'en\'erateur de conditions de v\'erification et un analyseur statique
qui peuvent \^etre utilis\'es dans des preuves par r\'eflexion.  L'extraction
d'un interpr\`ete \`a partir de la s\'emantique d\'enotationnelle est \'egalement
d\'ecrite.  Des preuves formelles assurant la correction des diff\'erents
aspects vis-\`a-vis de la s\'emantique naturelle sont \'egalement abordées.} 
\RRkeyword{Coq, natural semantics, structural operational semantics,
denotational semantics,
axiomatic semantics, abstract interpretation, formal verification,
calculus of inductive constructions, proof by reflection, program
extraction}
\RRmotcle{Coq, s\'emantique naturelle, s\'emantique d\'enotationnelle,
s\'emantique
axiomatique, interpr\'etation abstraite, v\'erification
formelle, calcul des constructions inductives, preuve par r\'eflexion,
extraction de programmes}
\RRprojet{Marelle}
\RRtheme{\THSym}
\URSophia
\usepackage{alltt}
\usepackage{amssymb}
\newcommand{\coqand}{\(\wedge\)}
\renewcommand{\u}{{\tt\_}}

\begin{document}
\makeRR{\em This paper is dedicated to the memory of Gilles Kahn, my thesis advisor,
my mentor, my friend.}

\section{introduction}
Nipkow demonstrated in \cite{DBLP:journals/fac/Nipkow98} that theorem
provers could be used to formalize many aspects of programming
language semantics.  In this paper, we want to push the experiment
further to show that this formalization effort also has a practical
outcome, in that it makes it possible to integrate programming tools
inside theorem provers in an uniform way.  We re-visit the study of
operational, denotational semantics, axiomatic semantics, and weakest
pre-condiction calculus as already studied by Nipkow and we add a
small example of a static analysis tool based on abstract
interpretation.

To integrate the programming tools
inside the theorem prover we rely on the possibility to execute the
algorithms after they have been formally described inside the theorem
prover and to use theorems about these algorithms to assert properties
of the algorithm's input, a technique known as {\em reflection}
\cite{BartheRuysBarendregt95,Boutin97b}.  Actually, we also implemented a
parser, so that the theorem prover can be used as a playground to
experiment on sample programs.  We performed this experiment using the
Coq system \cite{Coq,Coqart}.  The tools that are formally described
can also be ``extracted'' outside the proof environment, so that they
become stand alone programs, thanks to the extracting capabilities
provided in \cite{Letouzey2002types}.

The desire to use computers to verify proofs about programming language
semantics was probably one of the main incentives for the design of modern
interactive theorem provers.  The LCF system was a pioneer in this
direction.  The theory of programming
languages was so grounded in basic mathematics that a tool like LCF
was quickly recognized as a tool in which mathematical reasoning can
also be simulated and proofs can be verified by decomposing them in
sound basic logical steps.  LCF started a large family of theorem
proving tools, among which HOL \cite{HOL} and Isabelle \cite{Paulson:Isabelle}
have achieved an outstanding international recognition.  Nipkow's experiments
were conducted using Isabelle.

In the family of theorem proving tools, there are two large
sub-families: there are the direct descendants of the LCF system
\cite{DBLP:books/sp/Gordon79},
which rely on simply-typed \(\lambda\)-calculus and the axioms of
higher-order logic to provide foundations for a large portion of
mathematics; on the other hand, there are systems descending from
de~Bruijn's Automath system and Martin-L{\"o}f's theory of types,
where propositions are directly represented as types,
``non-simple'' types, namely dependent types, can be used to represent
quantified statements, and typed functions are directly used to
represent proofs (the statement they prove being their type).
In systems of the LCF family, typed \(\lambda\)-terms are used in the
representation of logical statements and proofs are objects of another
nature.  In systems of the latter family, usually called type
theory-based theorem proving tools, typed \(\lambda\)-terms are used
both in the representation of logical statements and in the
representation of proofs.  Well-known members of the type theory based
family of theorem proving tools are Nuprl \cite{NUPRL}, Agda \cite{Agda},
and Coq.

The fact that typed \(\lambda\)-terms are used both to represent
logical statements and proofs in type theory-based theorem proving
tool has the consequence that computation in the typed
\(\lambda\)-calculus plays a central role in type-theory based theorem
proving tools, because verifying that a theorem is applied to an
argument of the right form may require an arbitrary large computation
in these systems.  By contrast, computation plays only a secondary
role in LCF-style theorem proving tools and facilities to execute programs
efficiently inside the theorem prover to support proofs was only added
recently \cite{DBLP:conf/types/BerghoferN00}.

With structural operational semantics and natural semantics, Gordon
Plotkin and Gilles Kahn provided systematic approaches to describing
programming languages relying mostly on the basic concepts of
inductive types and inductive propositions.  Execution states are
represented as environments, in other words lists of pairs binding a
variable name and a value.  Programs themselves can also be
represented as an inductive data-type, following the tradition of {\em
  abstract syntax} trees, a streamlined form of parsing trees.
Execution of instructions can then be described as inductive
propositions, where executing an instruction is described as a ternary
relation between an input environment, an instruction, and an output
value.  The execution of each program construct is described by
composing ``smaller'' executions of this construct or its
sub-components.  We will show that descriptions of execution can also be
represented using functions inside the theorem prover and we will prove
that these functions are consistent with the initial semantics, in effect
producing certified interpreters for the studied language.

Another approach to describing the behavior of programs is to express
that a program links properties of inputs with properties of outputs.
In other words, one provide a logical system to describe under which
condition on a program's input a given condition on the program's
output can be guaranteed (as long as the program terminates).  This
style of description is known as {\em axiomatic semantics} and was
proposed by Hoare \cite{hoare69}.  Here again, we can
use an inductive type to represent a basic language of properties of
input and output of programs.  We will show that axiomatic semantics can easily
be described using inductive properties and recursive functions and again
we will show that the new reasoning rules are consistent with the initial
operational semantics.  Axiomatic semantics also support an algorithmic
presentation, known as a {\em verification condition generator}
for the {\em weakest pre-condition calculus} as
advocated by Dijkstra \cite{Dijkstra76}.  Again, we provide an
implementation of this generator and a proof that it is correct.
Thanks to the reflection approach, this generator can be used inside the
theorem prover to establish properties of sample programs.

The next style of semantic description for programming language that
we will study will be the style known as {\em denotational semantics}
or {\em domain theory}, actually the style that motivated the first
implementation of the LCF system.  Here, the semantics of the
instructions is described as a collection of partial functions from a
type of inputs to a type of outputs.  The kind of functions that are
commonly used in type-theory based theorem proving tools are not
directly suited for this approach, for fundamental reasons.  We will show
what axioms of classical logical can be used to provide a simple encoding
of the partial functions we need.  However, using these axioms precludes
computing inside the theorem prover, so that the function we obtain are
executable only after extraction outside the theorem prover.  This approach
can still be used to derive an interpreter, a tool to execute programs,
with a guarantee that the interpreter respects the reference operational
semantics.

The last category of semantic approaches to programming languages that we
want to address in this paper is an approach to the static analysis of
programs known as {\em abstract interpretation}.  While other approaches
aim at giving a completely precise understanding of what happens in programs,
abstract interpretation focusses on providing abstract views of execution.
The goal is to hide enough details so that the information that is obtained
from the analysis is easier to manage and more importantly the computations
to perform the analysis can be performed by a program that is guaranteed
to terminate.

\subsection{Related work}
The main reference we used on programming language semantics is Winskel's
text book \cite{Winskel93}.

Many publications have been provided to show that these various aspects
of programming language could be handled in theorem provers.  Our first
example is \cite{reasoningexecutable} where we described the correctness
of a program transformation tool with respect to the language's operational
semantics.  This work was performed in the context of the Centaur system
\cite{CentaurShort} where semantic descriptions could be executed with the
help of a prolog interpreter or reasoned about using a translation to
the Coq theorem prover \cite{Ter95short}.  The most impressive experiment
is described in \cite{DBLP:journals/fac/Nipkow98}, who approximately
formalizes the first
100 pages of Winskel's book, thus including a few more proofs around the
relations between operational semantics, axiomatic semantics, and denotational
semantics than we describe here. The difference between our work
and Nipkow's is that we rely more on reflection and make a few different
choices, like the choice to provide a minimal syntax for assertions, while
Nipkow directly uses meta-level logical formulas and thus avoid the need
to describe substitution.  On the other hand, our choice of an abstract
syntax for assertions makes it possible to integrate our verification generator
with a parser, thus providing a more user-friendly approach to annotated
programs.

The work on denotational semantics is a transposition and a reduction
of the work on domain theory that could already
be described formally in the framework
of {\em logic of computable functions}, in Isabelle
\cite{MuellerNvOS99}.

The study of interactions between abstract interpretation and theorem provers
is the object of more recent work.  Intermediate approaches use abstract
interpreters to generated proofs of correctness of programs in axiomatics
semantics as in \cite{DBLP:conf/ictac/Chaieb06}.  Pichardie \cite{PichardieThesis}
actually goes all the way to formally describing a general framework
for abstract interpretation and then
instantiating it for specific domains to obtain static analysis tools. 
Our work is similar except that Pichardie's work is based on transition
semantics, this imposes that recursion is based 
on well-founded recursion, a feature that makes it ill-suited for use in
reflection.

Application domains for theorem prover-aware formal semantics of programming
languages abound.  Nipkow and his team \cite{Oheimb2000}, Jacobs and his team,
\cite{LOOPCompiler}, and Barthe and his team \cite{g+01:esop,Bertot00a} showed
the benefits there could be in describing the Java programming language and
the Java virtual machine, to verify soundness properties of the byte-code
verifier and apply this the guarantees of the security that the Java language
and its Smartcard-aware offspring, JavaCard.  More recent work by Leroy and
his team show that this work can be extended to the formalization of efficient
compilers.
\section{Concrete and abstract syntax}
We consider a {\em while loop} programing language with
simple arithmetic expressions: it is the {\sf Imp} language of
\cite{DBLP:journals/fac/Nipkow98} without the conditional instruction.
The language has been trimmed to a
bare skeleton, but still retains the property of being Turing complete.
We will use \(\rho\) as meta-variables for variable declarations (we
will also often use the word {\em environment}), \(e\) for expressions,
\(b\) for boolean expressions, and \(i\) for instructions.  We use
an infinite denumerable set of variable names whose elements are
written \(x, y, x_1, \ldots\) and we use \(n, n_1, n'\) to represent integers.
The syntactic categories are defined as follows:
\[\rho ::= (x,n)\cdot\rho | \emptyset \qquad e ::= n~ |~ x~ |~ e {\sf+}e \qquad
b::= e{\tt<}e\]
\[i::= {\sf skip} ~|~ x{\sf:=}e~|~ i{\sf;}i ~|~{\sf while}~b~{\sf do}~i~{\sf done}\]
The intended meaning of most of these constructs should be obvious.  The only
suprising element may be the {\sf skip} instruction: this is an {\em empty}
program, which does nothing.

In the theorem prover, we use inductive types to describe these
syntactic categories.  The convention that numbers are expressions needs
to be modified: there is a constructor {\sf anum} in the type 
of arithmetic expression {\sf aexpr} that maps
a number to the corresponding expression.  Similarly, variable names
are transformed into arithmetic expressions and assignments just use
variable names as first components.
\begin{alltt}\sf
Inductive aexpr : Type := avar (s : string) \(|\) anum (n : Z)  \(|\) aplus (a\(\sb{1}\) a\(\sb{2}\) :aexpr).

Inductive bexpr : Type := blt (a\(\sb{1}\) a\(\sb{2}\) : aexpr).

Inductive instr : Type :=
  assign (s: string)(e:aexpr) \(|\) sequence (i\(\sb{1}\) i\(\sb{2}\):instr) \(|\) while (b:bexpr)(i:instr) \(|\) skip.
\end{alltt}
\section{Operational semantics}
\subsection{Evaluation and environment update}
\subsubsection{Inference rules}
We will describe the evaluation of expressions using judgments of the
form \hbox{\(\rho\vdash e\rightarrow v\)} or 
\hbox{\(\rho\vdash b\rightarrow v\)} (with
a straight arrow).  These judgments should be read as {\em in
environment \(\rho\), the arithmetic expression \(e\)
(resp. the expression \(b\)) has the value \(v\).}  The value \(v\) is an
integer or a boolean value depending on the kind of expression being evaluated.
The rules describing evaluation are as follows:
\[{\frac{\displaystyle \strut}{\displaystyle\rho\vdash n\rightarrow n}}\qquad
{\frac{\displaystyle \strut}{\displaystyle(x,n)\cdot\rho\vdash x\rightarrow n}}\]
\[{\frac{\displaystyle \rho\vdash x\rightarrow n\quad x\neq y}{\displaystyle(y,n')\cdot\rho\vdash x\rightarrow n}}\qquad
{\frac{\displaystyle \rho\vdash e_1\rightarrow n_1\quad\rho\vdash e_2\rightarrow n_2}{\displaystyle\rho\vdash e_1{\sf+}e_2\rightarrow n_1+n_2}}\]
\[{\frac{\displaystyle \rho\vdash e_1\rightarrow n_1\quad\rho\vdash e_2\rightarrow n_2\quad n_1 < n_2}
{\displaystyle\rho\vdash e_1{\sf<}e_2\rightarrow {\sf true}}}\qquad
{\frac{\displaystyle \rho\vdash e_1\rightarrow n_1\quad\rho\vdash e_2\rightarrow n_2\quad n_2 \leq n_1}{\displaystyle\rho\vdash e_1{\sf<}e_2\rightarrow {\sf false}}}\]
During the execution of instructions, we will regularly need describing
the modification of an environment, so that the value associated to a variable
is modified.  We use judgments of the form \(\rho \vdash x, n\mapsto \rho'\),
which should be read as {\em \(x\) has a value in \(\rho\) and \(\rho'\) and
the value for \(x\) in \(\rho'\) is \(n\); every other variable that has
a value in \(\rho\) has the same value in \(\rho'\).}
 This is simply described using two inference rules, in the same spirit as
rules to evaluate variables.
\subsubsection{Theorem prover encoding}
Judgments of the form \(\cdot\vdash \cdot\rightarrow \cdot\) are represented
by three-argument inductive predicates named {\sf aeval} and {\sf beval}.
We need to have two predicates to account for the fact that the same judgment
is actually used to describe the evaluations of expressions of two different
types.  The encoding of premises is quite straight forward using nested
implications, and we add universal quantifications for every variable that
occurs in the inference rules.  All inference rules for a given judgment
are grouped
in a single inductive definition.  This makes it possible to express that
the meaning of the judgment \(\cdot\vdash \cdot\rightarrow\cdot\) is
expressed by these inferences {\em and only these inferences rules}.  

Environments are encoded as lists of pairs of a string and an
integer, so that the environment \(\emptyset\) is encoded as {\sf nil}
and the environment \((x,n)\cdot\rho\) is {\sf (x,n)::r}.

\begin{alltt}\sf
Definition env := list(string*Z).

Inductive aeval : env \(\rightarrow\) aexpr \(\rightarrow\) Z \(\rightarrow\) Prop :=
  ae\u{}int : \(\forall\) r n, aeval r (anum n) n
\(|\) ae\u{}var\(\sb{1}\) : \(\forall\) r x n, aeval ((x,n)::r) (avar x) n
\(|\) ae\u{}var\(\sb{2}\) : \(\forall\) r x y v v' , x \(\neq\) y \(\rightarrow\) aeval r (avar x) v \(\rightarrow\) aeval ((y,v')::r) (avar x) v
\(|\) ae\u{}plus : \(\forall\) r e\(\sb{1}\) e\(\sb{2}\) v\(\sb{1}\) v\(\sb{2}\), aeval r e\(\sb{1}\) v\(\sb{1}\) \(\rightarrow\) aeval r e\(\sb{2}\) v\(\sb{2}\) \(\rightarrow\)
              aeval r (aplus e\(\sb{1}\) e\(\sb{2}\)) (v\(\sb{1}\) + v\(\sb{2}\)).

Inductive beval : env \(\rightarrow\) bexpr \(\rightarrow\) bool \(\rightarrow\) Prop :=
\(|\) be\u{}lt\(\sb{1}\) : \(\forall\) r e\(\sb{1}\) e\(\sb{2}\) v\(\sb{1}\) v\(\sb{2}\), aeval r e\(\sb{1}\) v\(\sb{1}\) \(\rightarrow\) aeval r e\(\sb{2}\) v\(\sb{2}\) \(\rightarrow\) v\(\sb{1}\) \(<\) v\(\sb{2}\) \(\rightarrow\)
            beval r (blt e\(\sb{1}\) e\(\sb{2}\)) true
\(|\) be\u{}lt\(\sb{2}\) : \(\forall\) r e\(\sb{1}\) e\(\sb{2}\) v\(\sb{1}\) v\(\sb{2}\), aeval r e\(\sb{1}\) v\(\sb{1}\) \(\rightarrow\) aeval r e\(\sb{2}\) v\(\sb{2}\) \(\rightarrow\) v\(\sb{2}\) \(\leq\) v\(\sb{1}\) \(\rightarrow\)
            beval r (blt e\(\sb{1}\) e\(\sb{2}\)) false.
\end{alltt}
the four place judgment \(\cdot\vdash\cdot,\cdot\mapsto\cdot\) is also encoded
as an inductive definition for a predicate named {\sf update}.

Induction principles are automatically generated for these
declarations of inductive predicates.  These induction principles are
instrumental for the proofs presented later in the paper.
\subsection{Functional encoding}
The judgment \(\rho\vdash e\rightarrow n\) actually describes a partial
function: for given \(\rho\) and \(e\), there is at
most one value \(n\) such that \(\rho\vdash e \rightarrow n\) holds.
We describe this function in two steps with {\sf lookup}
and {\sf af}, which return values in {\sf option Z}.  When computing
additions, we need to compose total functions with partial
functions. For this, we define a {\sf bind} function that takes
care of undefined values in intermediate results.  The pre-defined
function {\sf string\u{}dec} is used to compare two strings.
\begin{alltt}\sf
Fixpoint lookup (r:env)(s:string)\{struct r\} : option Z :=
match r with
 nil \(\Rightarrow\) None  \(|\)  (a,b)::tl \(\Rightarrow\) if (string\u{}dec a s) then Some b else lookup tl s
end.

Definition bind (A B:Type)(v:option A)(f:A\(\rightarrow\)option B) : option B :=
  match v with Some x \(\Rightarrow\) f x \(|\) None \(\Rightarrow\) None end.

Fixpoint af (r:env)(a:aexpr) : option Z :=
match a with
  avar index \(\Rightarrow\) lookup r index \hskip1cm \(|\) anum n \(\Rightarrow\) Some n 
\(|\) aplus e\(\sb{1}\) e\(\sb{2}\) \(\Rightarrow\) bind (af r e\(\sb{1}\)) (fun v\(\sb{1}\) \(\Rightarrow\) bind (af r e\(\sb{2}\)) (fun v\(\sb{2}\) \(\Rightarrow\) Some (v\(\sb{1}\)+v\(\sb{2}\))))
end.
\end{alltt}
We can define functions {\sf bf} to evaluate boolean expressions and 
{\sf uf} to compute updated environments in a similar way.

We use two functions to describe the evaluation of arithmetic expressions,
because evaluating variables requires a recursion where the environment
decreases at each recursive call (the expression staying fixed), while
the evaluation of additions requires a recursion where the expression
decreases at each recursive call (the environment staying fixed).  The
{\sf Fixpoint} construct imposes that the two kinds of recursion should be
separated.

With {\sf aeval} and {\sf af}, we have two encodings of the same
concept.  We need to show that these encoding are equivalent, this
is done with the following lemmas.
\begin{alltt}\sf
Lemma lookup\u{}aeval : \(\forall\) r s v, lookup r s = Some v \(\rightarrow\) aeval r (avar s) v.

Lemma af\u{}eval : \(\forall\) r e v, af r e = Some v \(\rightarrow\) aeval r e v.

Lemma aeval\u{}f : \(\forall\) r e n, aeval r e n \(\rightarrow\) af r e = Some n.
\end{alltt}
The proof of the first lemma is done by induction on the structure of {\sf r},
the proof of the second lemma is done by induction on {\sf e}, while
the proof of the third lemma is done by induction on the structure of the
proof for {\sf aeval} (using the induction principle, which is generated
when the inductive predicate is declared).  Using simple proof commands,
each of these proofs is less than ten lines long.  We also have similar
correctness proofs for {\sf bf} and {\sf uf}.
\subsection{Natural semantics}
With {\em natural semantics} \cite{natural}, Gilles Kahn proposed that
one should rely on judgments expressing the execution of program fragments
until they terminate.  The same style was also called {\em big-step}
semantics.  The main advantage of this description style is that it supports
very concise descriptions for sequential languages.  For our little language
with four instructions, we only need five inference rules.

We rely on judgments of the form \(\rho \vdash i \leadsto \rho'\) (with
a twisted arrow).  These judgments should be read as {\em executing
\(i\) from the initial environment \(\rho\) terminates and yields the
new environment \(\rho'\)}.
\[\overline{\rho\vdash {\sf skip} \leadsto \rho}\qquad
{\frac{\displaystyle \rho\vdash e \rightarrow n\qquad
\rho\vdash x,n\mapsto \rho'}
{\displaystyle \rho\vdash x {\sf:=} e \leadsto \rho'}}\]
\[{\frac{\displaystyle \rho\vdash i_1 \leadsto \rho' \qquad
\rho'\vdash i_2\leadsto \rho''}
{\displaystyle \rho\vdash i_1{\sf;}i_2\leadsto \rho''}}\qquad
{\frac{\displaystyle \rho\vdash b\rightarrow {\sf false}}
{\displaystyle \rho\vdash{\sf while}~b~{\sf do}~i~{\sf done}\leadsto \rho}}\]
\[{\frac{\displaystyle \rho\vdash b\rightarrow {\sf true}\quad
\rho\vdash i\leadsto \rho'\quad
\rho'\vdash {\sf while}~b~{\sf do}~i~{\sf done}\leadsto \rho''}
{\displaystyle \rho\vdash {\sf while}~b~{\sf do}~i~{\sf done}\leadsto \rho''}}
\]
Because it is described using collections of rules, the judgment
\(\cdot\vdash \cdot\leadsto \cdot\) can be described
with an inductive predicate exactly like the judgments for
evaluation and update.  We use the name {\sf exec} for this judgment.

Like the judgment \(\rho\vdash e\rightarrow v\), the judgment
\(\rho\vdash i \leadsto \rho'\) actually describes a partial function.  However,
this partial function cannot be described as a structural recursive function as 
we did when defining the functions {\sf lookup} and {\sf af}.  For
while loop, Such a function
would present a recursive call where neither
the environment nor the instruction argument would be a sub-structure of the
corresponding initial argument.  This failure also relates to the fact that
the termination of programs is undecidable for this kind of language, while
structural recursion would provide a terminating tool to compute whether
programs terminate.  In the later section on denotational semantics, we will
discuss ways to encode
a form of recursion that is powerful enough to describe the semantics as
a recursive function.
\section{Axiomatic semantics}
We study now the encoding of axiomatic semantics as proposed by
Hoare \cite{hoare69} and the weakest pre-condition calculus as
proposed by Dijkstra \cite{Dijkstra76}.
The principle of this semantic approach is to consider properties that
are satisfied by the variables of the program before and after the
execution.

\subsection{The semantic rules}
To describe this approach, we use judgments of the following form:
\(\{P\} i \{Q\}\).  This should be read as {\em if \(P\) is satisfied
before executing \(i\) and executing \(i\) terminates, then \(Q\) is
guaranteed to be satisfied after executing \(i\)}.

There are two key aspects in axiomatic semantics: first the behavior
of assignment is explained by substituting variables with
arithmetic expressions; second the behavior of control operators is
explained by isolating properties that are independent from the
choice made in the control operator and properties that can be
deduced from the choice made in the control operator.
\[\overline{\{P\}{\sf skip}\{P\}}\qquad
{\frac{\displaystyle\{P\}i_1\{Q\}\quad\{Q\}i_2\{R\}}
{\displaystyle\{P\}i_1{\sf ;}i_2\{R\}}}\]
\[{\overline{\{P[x\leftarrow e]\}x{\sf:=}e\{P\}}}
\qquad{\frac{\displaystyle\{b\wedge P\}i\{P\}}
{\displaystyle\{P\}{\sf while}~b~{\sf do}~i~{\sf done}\{\neg b\wedge P\}}}
\]
\[{\frac{P\Rightarrow P_1\quad\{P_1\}i\{Q_1\}\quad Q_1\Rightarrow Q}
{\{P\}i\{Q\}}}\]
In the rule for while loops, the property \(P\) corresponds to
something that should be verified whether the loop body is executed 0,
1, or many times: it is independent from the choice made in the
control operator.  However, when the loop terminates, one knows that
the test must have failed, this is why the output property for the
loop contains \(\neg b\).  Also, if \(P\) should be preserved
independently of the number of executions of the loop, then it should
be preserved through execution of the loop body, but only when
the test is satisfied.

We call the first four rules {\em structural rules}: each of them
handles a construct of the programming language.
The last rule, known as the {\em consequence} rule, makes it
possible to mix logical reasoning about the properties
with the symbolic reasoning about the program constructs.  To prove
the two premises that are implications, it is necessary to master the
actual meaning of the properties, conjunction, and negation.
\subsection{Theorem prover encoding}
The first step is to define a data-type for assertions.
Again, we keep things minimal.  Obviously, the inference rules require that
the language of assertions contain at least conjunctions, negations, 
and tests from the language's boolean expressions.
We also include the possibility to have abitrary predicates on arithmetic
expressions, represented by a name given as a string.
\begin{alltt}\sf
Inductive assert : Type :=
  a\u{}b (b: bexpr) \(|\) a\u{}not (a: assert) \(|\) a\u{}conj (a a': assert) \(|\) pred (s: string)(l: int aexpr).

Inductive condition : Type := c\u{}imp (a a':assert).
\end{alltt}

For variables
that occur inside arithmetic expressions, we use valuation functions of
type {\sf string\(\rightarrow\) Z} instead of environments and we define
a new function {\sf af'} (respectively {\sf bf'}, {\sf lf'}) to
compute the value
of an arithmetic expression (respectively boolean expressions, lists
of arithmetic expressions)
for a given valuation.
The function {\sf af'} is more
practical to use and define than {\sf af} because it is total, while {\sf af} was partial.
\begin{alltt}\sf
Fixpoint af' (g:string\(\rightarrow\)Z)(a:aexpr) : Z :=
match a with avar s \(\Rightarrow\) g s \(|\) anum n \(\Rightarrow\) n \(|\) aplus e\(\sb{1}\) e\(\sb{2}\) \(\Rightarrow\) af' g e\(\sb{1}\) + af' g e\(\sb{2}\) end.
\end{alltt}
To give a meaning to predicates, we use lists of pairs associating names
and predicates on lists of integers as {\em predicate} environments and we
have a function {\sf f\u{}p} to map an environment and a string to a
predicate on integers.  

With all these functions, we can interpret assertions as propositional
values using a function {\sf i\u{}a} and conditions using a function
{\sf i\u{}c}.
\begin{alltt}\sf
Definition p\u{}env := list(string*(list Z\(\rightarrow\)Prop)).

Fixpoint i\u{}a (m: p\u{}env)(g:string\(\rightarrow\)Z)(a:assert) : Prop :=
  match a with
    a\u{}b e \(\Rightarrow\) bf' g e\hskip46pt  \(|\) a\u{}not a \(\Rightarrow\) ~ i\u{}a m g a
  \(|\) pred p l \(\Rightarrow\) f\u{}p m p (lf' g l)  \(|\) a\u{}conj a\(\sb{1}\) a\(\sb{2}\) \(\Rightarrow\) i\u{}a m g a\(\sb{1}\) \coqand{} i\u{}a m g a\(\sb{2}\)
  end.

Definition i\u{}c (m:p\u{}env)(g:string\(\rightarrow\)Z)(c:condition) :=
 match c with c\u{}imp a\(\sb{1}\) a\(\sb{2}\) \(\Rightarrow\) i\u{}a m g a\(\sb{1}\) \(\rightarrow\) i\u{}a m g a\(\sb{2}\) end.
\end{alltt}
The validity of conditions can be expressed for a given predicate environment
by saying that their interpretation should hold for any valuation.
\begin{alltt}\sf
Definition valid (m:p\u{}env)(c:condition) := \(\forall\) g, i\u{}c m g c.
\end{alltt}
We also define substitution for arithmetic expressions, boolean expressions,
and so on, each time traversing structures.  The function 
at the level of assertions is called {\sf a\u{}subst}.  We can then define
the axiomatic semantics.
\begin{alltt}\sf
Inductive ax\u{}sem (m :p\u{}env): assert \(\rightarrow\) instr \(\rightarrow\) assert \(\rightarrow\) Prop:=
  ax\(\sb{1}\) : \(\forall\) P, ax\u{}sem m P skip P
\(|\) ax\(\sb{2}\) : \(\forall\) P x e, ax\u{}sem m (a\u{}subst P x e) (assign x e) P
\(|\) ax\(\sb{3}\) : \(\forall\) P Q R i\(\sb{1}\) i\(\sb{2}\), ax\u{}sem m P i\(\sb{1}\) Q \(\rightarrow\) ax\u{}sem m Q i\(\sb{2}\) R \(\rightarrow\)
          ax\u{}sem m P (sequence i\(\sb{1}\) i\(\sb{2}\)) R
\(|\) ax\(\sb{4}\) : \(\forall\) P b i, ax\u{}sem m (a\u{}conj (a\u{}b b) P) i P \(\rightarrow\)
          ax\u{}sem m P (while b i) (a\u{}conj (a\u{}not (a\u{}b b)) P)
\(|\) ax\(\sb{5}\) : \(\forall\) P P' Q' Q i,
        valid m (c\u{}imp P P') \(\rightarrow\) ax\u{}sem m P' i Q' \(\rightarrow\) valid m (c\u{}imp Q' Q) \(\rightarrow\)
        ax\u{}sem m P i Q.
\end{alltt}
\subsection{Proving the correctness}
We want to certify that the properties of programs that we can prove using
axiomatic semantics hold for actual executions of programs, as described
by the operational semantics.  We first define a mapping from
the environments used in operational semantics to the valuations used
in the axiomatic semantics.  This mapping is called {\sf e\u{}to\u{}f},
the expression {\sf e\u{}to\u{}f \(e\) \(g\) \(x\)} is
the value of \(x\) in the environment \(e\), when it is
defined, and \(g~x\) otherwise.  The formula 
{\sf e\u{}to\u{}f \(e\) \(g\)} is also written {\sf \(e\)@\(g\)}.
We express the correctness of axiomatic semantics
by stating that if ``{\sf exec} \(r\) \(i\) \(r'\)'' and
``{\sf ax\u{}sem \(P\) \(i\) \(Q\)}'' hold, if \(P\) holds in the initial
environment, \(Q\) should hold in the final environment \(Q\).
\begin{alltt}\sf
Theorem ax\u{}sem\u{}sound : \(\forall\) m r i r' g P Q, exec r i r' \(\rightarrow\) ax\u{}sem m P i Q \(\rightarrow\)
   i\u{}a m (r@g) P \(\rightarrow\) i\u{}a m (r'@g) Q.
\end{alltt}
When we attempt to prove this statement by induction on {\sf exec}
and case analyis on {\sf ax\u{}sem}, we encounter problem because uses
of consequence rules may make axiomatic semantics derivations arbitrary large.
To reduce this problem we introduce a notion of 
 {\em normalized} derivations where
exactly one consequence step is associated to every structural step.
We introduce an extra inductive predicate call {\sf nax} to model these
normalized derivation, with only four
constructors.  For instance, here is the constructor for loops:
\begin{alltt}\sf
  nax\(\sb{4}\) : \(\forall\) P P' Q b i,
    valid m (c\u{}imp P P') \(\rightarrow\) valid m (c\u{}imp (a\u{}conj (a\u{}not (a\u{}b b)) P') Q) \(\rightarrow\)
    nax m (a\u{}conj (a\u{}b b) P') i P' \(\rightarrow\) nax m P (while b i) Q.
\end{alltt}
We prove that {\sf ax\u{}sem} and {\sf nax} are equivalent.
This ``organisational'' step is crucial.  We can now prove the correctness
statement by induction on {\sf exec} and by cases on {\sf nax},
while a proof by double induction would be required with {\sf ax\u{}sem}.

Another key lemma shows that updating an environment for a variable
and a value, as performed in operational semantics, and substituting
an arithmetic expression for a variable, as performed in axiomatic
semantics, are consistent.
\begin{alltt}\sf
Lemma a\u{}subst\u{}correct : forall a r1 e v m g r2 x,
    aeval r1 e v \(\rightarrow\) s\u{}update r1 x v r2 \(\rightarrow\)
    (i\u{}a m (r1@g) (a\u{}subst a x e) \(\leftrightarrow\) i\u{}a m (r2@g) a).
\end{alltt}
\subsection{The weakest pre-condition calculus}
Most of the
structure of an axiomatic semantics proof can be deduced from
the structure of the instruction.
However, the assertions in loop invariants and in consequence
rules cannot be guessed.  Dijkstra proposed to annotate programs with the
unguessable formulas and to automatically gather the
implications used in consequence steps as a collection of conditions
to be proved on the side.  The result is a {\em verification
condition generator} which takes annotated program as input and returns
a list of conditions.  We will now show how to encode such a 
verification condition generator ({\sf vcg}).

We need to define a new data-type for these annotated programs.
\begin{alltt}\sf
Inductive a\u{}instr : Type :=
  prec (a:assert)(i:a\u{}instr) \(|\) a\u{}skip \(|\) a\u{}assign (s:string)(e:aexpr)
\(|\) a\u{}sequence (i\(\sb{1}\) i\(\sb{2}\):a\u{}instr) \(|\) a\u{}while (b:bexpr)(a:assert)(i:a\u{}instr).
\end{alltt}
The {\sf prec} constructor is used to assert
properties of a program's variables at any point in the program.

The computation of all the implications works in two steps.
The first step is to understand what is the pre-condition for
an annotated instruction and a given post-condition.  For the
{\sf a\u{}while} and {\sf prec} constructs, the pre-condition is simply
the one declared in the corresponding annotation, for the other constructs,
the pre-condition has to be computed using substitution and composition.
\begin{alltt}\sf
Fixpoint pc (i:a\u{}instr)(a:assert) \{struct i\} : assert :=
  match i with
    prec a' i \(\Rightarrow\) a'    \(|\) a\u{}while b a' i \(\Rightarrow\) a'   \(|\) a\u{}skip \(\Rightarrow\) a
  \(|\) a\u{}assign x e \(\Rightarrow\) a\u{}subst a x e              \(|\) a\u{}sequence i\(\sb{1}\) i\(\sb{2}\) \(\Rightarrow\) pc i\(\sb{1}\) (pc i\(\sb{2}\) a)
  end.
\end{alltt}
The second step is to gather all the conditions that would appear
in a minimal axiomatic semantics proof for the given post-condition,
starting from the corresponding pre-condition.
\begin{alltt}\sf
Fixpoint vcg (i:a\u{}instr)(post : assert) \{struct i\} : list condition :=
match i with
  a\u{}skip \(\Rightarrow\) nil     \(|\) a\u{}assign \u{} \u{} \(\Rightarrow\) nil   \(|\) prec a i \(\Rightarrow\) c\u{}imp a (pc i post)::vcg i post
\(|\) a\u{}sequence i\(\sb{1}\) i\(\sb{2}\) \(\Rightarrow\) vcg i\(\sb{2}\) post ++ vcg i\(\sb{1}\) (pc i\(\sb{2}\) post)
\(|\) a\u{}while e a i \(\Rightarrow\)
   c\u{}imp (a\u{}conj (a\u{}not (a\u{}b e)) a) post :: c\u{}imp (a\u{}conj (a\u{}b e) a) (pc i a) :: vcg i a
end.
\end{alltt}
The correctness of this verification condition generator is expressed
by showing that it suffices to prove the validity of all the
generated conditions to ensure that the Hoare triple holds.  This proof
is done by induction on the instruction.  We can then
obtain a proof that relates the condition generator and the operational
semantics.  In this statement, {\sf un\u{}annot} maps an annotated instruction
to the corresponding bare instruction.
\begin{alltt}\sf
Theorem vcg\u{}sound :
  \(\forall\) m i A, (valid\u{}l m (vcg i A)) \(\rightarrow\) \(\forall\) g r\(\sb{1}\) r\(\sb{2}\), exec r\(\sb{1}\) (un\u{}annot i) r\(\sb{2}\) \(\rightarrow\)
  i\u{}a m (e\u{}to\u{}f g r\(\sb{1}\)) (pc i A) \(\rightarrow\) i\u{}a m (e\u{}to\u{}f g r\(\sb{2}\)) A.
\end{alltt}
\subsection{An example of use in proof by reflection}
We consider the program that adds the \(n\) first positive integers.
We use 
a predicate environment {\sf ex\u{}m} that maps
two names {\sf le} and {\sf pp} to two predicates on lists
of two integers.  For the two integers \(x\) and \(y\), the
first predicate holds when \(x\leq y\) and the second holds when
\(2\times{}y=x\times(x+1)\).
With the help of a parser function, 
we can state the properties of interest for
our program in a concise manner:
\begin{alltt}\sf
Example ex\(\sb{1}\) : \(\forall\) g r\(\sb{2}\), 0 {\tt<} n \(\rightarrow\)
 exec (("x", 0)::("y", 0)::("n",n)::nil)
    (un\u{}annot (parse\u{}instr'
      "while x {\tt<} n do [le(x,n) \coqand{} pp(y,x)] x:=x+1;y:=x+y done")) r\(\sb{2}\) \(\rightarrow\) 
    2*(r\(\sb{2}\)@g)"y" = (r\(\sb{2}\)@g)"x"*((r\(\sb{2}\)@g)"x"+1).
\end{alltt}
After a few logistic steps, we can show that the conclusion is an
instance of the {\sf pp} predicate, and then apply the correctness
theorem, which leads to two logical requirements.  The first is
that the verification conditions hold:
\begin{alltt}\sf
valid\u{}l ex\u{}m
  (vcg (parse\u{}instr' "while x {\tt<} n do [le(x,n) \coqand{} pp(y,x)] x:=x+1;y:=x+y done")
    (parse\u{}assert' "pp(y,n)"))
\end{alltt}
After forcing the computation of the parser and the condition generator and
a few more logistic steps, this reduces to the following logical statement
\[
\begin{array}{l}
\forall x~ y~ n.\\
(x\not< n\wedge x\leq n \wedge 2y=x(x+1)\Rightarrow 2*y=n(n+1))\wedge\\
(x<n \wedge x\leq n\wedge 2y=x(x+1)\Rightarrow
x+1\leq n\wedge 2(x+1+y)=(x+1)(x+2)).
\end{array}
\]
This is easily proved using the regular Coq tactics.  The second requirement
is that the pre-condition should be satisfied and reduces to the statement
\[0\leq n\wedge 0=0.\]
We have actually automated proofs about programs inside the Coq system, thus
providing a simple model of tools like Why \cite{filliatre98}.
\section{Denotational semantics}
In denotational semantics, the aim is to describe the meaning of
instructions as functions.  The functions need to be partial, because
some instructions never terminate on some inputs.  We already used
partial functions for the functional encoding of expression evaluation.
However, the partial recursive function that we defined were structural,
and therefore guaranteed to terminate.  The execution function for instructions
does not fit in this framework and we will first define a new tool to
define recursive function.  Most notably, we will need to use non-constructive
logic for this purpose.

Again the partial functions will be implemented with the {\sf option}
inductive type, but the {\sf None} constructor will be used
to represent either
that an error occurs or that computation does not terminate.
\subsection{The fixpoint theorem}
The approach described in \cite{Winskel93} relies
on Tarski's fixpoint theorem, which states that every continuous
function in a complete partial order with a minimal element
has a least fixpoint and that this fixpoint is obtained by iterating
the function from the minimal element.

Our definition of complete partial order 
relies on the notion of chains, which are monotonic sequences.  A
partial order is a type with a relation \(\subseteq\) that is reflexive,
antisymmetric, and transitive; this partial order is complete if every
chain has a least upper bound.  A function \(f\) is continuous if for
every chain \(c\) with a least upper bound \(l\), the value \(f(l)\)
is the least upper bound of the sequence \(f(c_n)\).  Notice that when
defining continuous function in this way, we do not require \(f\) to
be monotonic; actually, we prove that every continuous function is
monotonic.

The proof of Tarski's theorem is quite easy to formalize and it can
be formalized using intuitionistic logic, so the plain calculus of
constructions is a satisfactory framework for this.
\subsection{Partial functions form a complete partial order}
The main work in applying Tarski's theorem revolves around proving
that types of partial functions are complete partial orders.
A type of the form {\sf option \(A\)} has the structure of a
complete partial order when choosing as order the relation such that
\(x\subseteq y\) exactly when \(x=y\) or \(x={\sf None}\).
The element {\sf None} is the minimal element.  Chains
have a finite co-domain, with at most two elements,
the least upper bound can be proved to exist using the non-consructive
excluded-middle axiom; this is our first step outside constructive
mathematics.

Given an arbitrary complete partial order \((B,\subseteq)\), the type
of functions of type \(A\rightarrow B\) is a complete partial order for
the order defined as follows:
\[f \subseteq g \Leftrightarrow \forall x, f(x)\subseteq g(x).\]
The proof that this is a complete partial order requires other
non-constructive axioms: extensionality is required to show that the order
is antisymetric and a description operator is required to construct
the least upper bound of a chain of functions.  We actually rely on
the non-constructive \(\epsilon\) operator proposed by Hilbert and already
used in HOL or Isabelle/HOL.  This \(\epsilon\) operator is a function that
takes
a type \(T\), a proof that \(T\) is inhabited, a predicate on \(T\), and 
returns a value in \(T\) that is guaranteed to satisfy the
predicate when possible.

For a sequence of functions \(f_n\) (not necessarily a chain), we can
define a new function \(f\), which maps every \(x\) to the value given
by the \(\epsilon\) operator for the predicate ``to be the least upper
bound of the sequence sequence \(f_n(x)\)''.  Now, if it happens that
\(f_n\) is a chain, then each of the sequences \(f_n(x)\) is a chain,
\(f(x)\) is guaranteed to be the least upper bound, and \(f\) is
the least upper bound of \(f_n\).

In practice, Tarski's least fixpoint theorem is a programming tool.
If one wishes to define a recursive function with a definition of
the form
\[f~x =  e\]
such that \(f\) appears in \(e\), it suffices that the function
\(\lambda f. \lambda x. e\) is continuous and the theorem returns a
function that satisfies this equation, a natural candidate for the
function that one wants to define.  We encode this fixpoint operator as
a function called {\sf Tarski\u{}fix}.
\subsection{Defining the semantics}
For a while loop of the form {\sf while}~\(b\)~{\sf do}~\(i\)~{\sf done},
such that the semantic function for \(i\) is \(f_i\),
we want the value of semantic function to be the function \(\phi_{b,i}\)
such that :
\[\phi_{b,i}(\rho) =
 \left\{\begin{array}{ll}{\sf \rho}&\hbox{if {\sf bf~\(b\)=false}}\\
\phi_{b,i}(\rho')~&\hbox{if {\sf bf~\(b\)=true} and \(f_i(\rho)={\sf Some}~\rho'\)}
\\
{\sf None}&\hbox{otherwise}\end{array}\right.\]
This function \(\phi_{b,i}\) is the least fixpoint
of the function {\sf F\u{}phi} obtained by combining a conditional construct,
a sequential composition function (already described using the {\sf bind}
function, and a few constant functions.
We encode {\sf F\u{}phi} and {\sf phi} as follows:
\begin{alltt}\sf
Definition ifthenelse (A:Type)(t:option bool)(v w: option A) :=
  match t with Some true \(\Rightarrow\) v \(|\) Some false \(\Rightarrow\) w \(|\) None \(\Rightarrow\) None end.

Notation "'IF x 'THEN a 'ELSE b" := (ifthenelse \u{} x a b) (at level 200). 

Definition F\u{}phi (A:Set)(t:A\(\rightarrow\)option bool)(f g :A\(\rightarrow\)option A) : A \(\rightarrow\) option A :=
   fun r \(\Rightarrow\) 'IF (t r) 'THEN (bind (f r) g) 'ELSE (Some r).
\end{alltt}
We proved that each of the constructs and {\sf F\u{}phi} are continuous.
The semantics for instructions can then be described by the following functions:
\begin{alltt}\sf
Definition phi := fun A t f \(\Rightarrow\) Tarski\u{}fix (F\u{}phi A t f).

Fixpoint ds(i:instr) : (list(string*Z)) \(\rightarrow\) option (list(string*Z)) :=
match i with
  assign x e \(\Rightarrow\) fun l \(\Rightarrow\) bind (af l e)(fun v \(\Rightarrow\) update l x v)
\(|\) sequence i\(\sb{1}\) i\(\sb{2}\) \(\Rightarrow\) fun r \(\Rightarrow\) (ds i\(\sb{1} r\))(ds i\(\sb{2}\))
\(|\) while e i \(\Rightarrow\) fun l \(\Rightarrow\) phi env (fun l' \(\Rightarrow\) bf l' e)(ds i) l
\(|\) skip \(\Rightarrow\) fun l \(\Rightarrow\) Some l
end.
\end{alltt}
We also proved the equivalence of this semantic definition and the
natural semantics specification:
\begin{alltt}\sf
Theorem ds\u{}eq\u{}sn : \(\forall\) i l l', ds i l = Some l' \(\leftrightarrow\) exec l i l'.
\end{alltt}
We actually rely on the second part of the fixpoint theorem,
which states that the least fixpoint of a continuous function is
the least upper bound of the chain obtained by iterating
the function on the least element.  In our case,
this gives the following corollary:
\[\forall x~v, \phi~x={\sf Some}~v\Rightarrow 
\exists n, {\sf F\u{}phi}^n~({\sf fun}~y\rightarrow {\sf None})~ x = {\sf Some}~x
\]
We can then proceed with a proof by induction on the number \(n\).

Unlike the functions {\sf af}, {\sf af'}, or {\sf vcg},
the function {\sf phi} is not usable
for computation inside the theorem prover, but {\sf F\u{}phi\(^n\)} can
be used to compute using approximations.
We can still extract this code and execute it in Ocaml, as long as we
extract the Tarski fixpoint theorem to a simple fixpoint function:
\begin{alltt}\sf
let rec fix f = f (fun y \(\rightarrow\) fix f y)
\end{alltt}
This interpreter loops when executing a looping program; this is predicted in
the Coq formalization by a value of {\sf None}.
\section{Abstract interpretation}
The goal of abstract interpretation \cite{CousotCousot77-1} is to
infer automatically properties about programs based on approximations
described as an abstract domain of values.  Approximations make it
possible to consider several executions at a time, for example all the
executions inside a loop.  This way the execution of arbitrary
programs can be approximated using an algorithm that has polynomial
complexity.

Abstract values are supposed to represent subsets of the set of
concrete values.  Each abstract interpreter works with a fixed set of
abstract values, which much have a certain structure.  An operation on
abstract values must be provided for each operation in the language
(in our case we only have to provide an addition on abstract values).
The subset represented by the result of an abstract operation must
contain all the values of the corresponding operation when applied to
values in the input subsets.  The set of abstract values should also
be ordered, in a way that is compatible with the inclusion order for
the subsets they represent.  Also, the type of abstract values should
also contain an element corresponding to the whole set of integers.
We will call this element the bottom abstract value.  The theoretical
foundations provided by Cousot and Cousot \cite{CousotCousot77-1}
actually enumerate all the properties that are required from the abstract
values.

Given an abstract valuation where variable names are mapped to abstract
values, we program an abstract evaluation function
{\sf ab\u{}eval} for arithmetic expressions that returns a new abstract
value.  This function is programmed exactly like the function {\sf af'}
we used for axiomatic semantics, simply replacing integer addition
with an abstract notion of addition on abstract values.

When we need to evaluate with respect to an abstract environment {\sf l}, i.e.,
a finite list of pairs of variable names and abstract values,
we use the function {\sf (ab\u{}lookup l)} that associates
the bottom value to all variables that do not occur in the abstract environment.

Abstract execution of instructions takes as input an abstract
environment and a bare instruction and returns the pair of an
annotated instruction and an optional final abstract environment.
When the optional environment is {\sf None}, this means that the
analysis detected that concrete execution never terminate.  The
annotations in the result instruction describe information that is
guaranteed to be satisfied when execution reaches the corresponding
point.

Abstract execution for assignments, sequences, and skip instructions
is natural to express: we just compute abtract values for expressions
and pass abstract environments around as we did in the concrete semantics.
For while loops, we handle them in a static way: our abstract
interpreter is designed as a tool that always terminate (even if the
analyzed program loops for ever).

The approach is to make the abstract environment coarser
and coarser, until we reach an approximation that is stable through abstract
interpretation of the loop body.  Thus, we want to find an invariant abstract
environment for loops, as we did in axiomatic semantics.
Finding the best possible approximation is undecidable and over-approximation
is required.  We chose to implement a simple strategy:
\begin{enumerate}
\item We first check whether the input abstract environment
for the while loop is {\em stable}, that is, if 
abstract values in the output
environment are included in the corresponding abstract value for the input
environment,
\item If this fails, we use a {\sf widen} function to compute an 
over-approximation of both the input and the output, and we then check whether
the new environment is stable,
{\sf widen}), and we check whether this new environment is stable,
\item If the first two steps failed, we overapproximate every value
with the bottom abstract value; this is necessarily stable but gives
no valuable information about any variable.
\end{enumerate}
We also incorporate information from the loop test.
When the test has the form \(v < e\), where \(v\) is a variable,
we can use this to refine the abstract value for \(v\).  At this
point, we may detect that the new abstract value
represents the empty set, this only happens when the test can never
succeed or never fail, and in this case some code behind this test is
dead-code.  This  is
performed by a function {\sf intersect\u{}env}.  This function takes
a first boolean argument that is used to express that we check whether
the test is satisfied or falsified.  This function returns {\sf None}
when the test can never be satisfied or can never be falsified.
When dead-code is detected, we mark the instruction with false assertions,
to express that the location is never reached, (this is
done in {\sf mark}).

To check for stability of environments,
we first need to combine the input and the output environment to find
a new environment where the value associated to each variable contains
the two values obtained from the two other environments.  This is done
with a function noted {\sf \(l\) @@ \(l'\)} (we named it {\sf join\u{}env}).  For a given while loop body, we call {\sf intersect\u{}env} and {\sf mark}
or {\sf @@} three times, one for every stage of our simple
strategy.  These operations are gathered in a function {\sf fp\(_1\)}.
\begin{alltt}\sf
Definition fp\(\sb{1}\)(l\(\sb{0}\) l:ab\u{}env)(b:bexpr)(i:instr)(f:ab\u{}env\(\rightarrow\) a\u{}instr*option ab\u{}env) :=
match intersect\u{}env true l b with
  None \(\Rightarrow\) (prec false\u{}assert (mark i), Some l)
\(|\) Some l' \(\Rightarrow\) let (i', l'') := f l' in
  match l'' with None \(\Rightarrow\) (i', None) \(|\) Some l\(\sb{2}\) \(\Rightarrow\) (i', Some (l\(\sb{0}\) @@ l' @@ l\(\sb{2}\))) end
end.
\end{alltt}
This function takes as argument the function {\sf f} that performs abstract
interpretation on the loop body {\sf i}.
We use this function {\sf fp\(\sb{1}\)} several times and combine it with 
widening functions to obtain a function {\sf fp} that performs our three
stage strategy.  When the result of {\sf fp} is {\sf (\(i\), Some \(l\))},
\(l\) satisfies the equation {\sf snd(f \(i\) \(l\))=\(l\)}.

Our abstract interpreter is then described
as a recursive function {\sf abstract\u{}i} (here we use {\sf to\u{}a} to
transform an environment into an assertion, and {\sf to\u{}a'} for optional
environments, mapping {\sf None} to {\sf false\u{}assert}).
\begin{alltt}\sf
Fixpoint abstract\u{}i (i : instr)(l : ab\u{}env) : a\u{}instr*option ab\u{}env :=
match i with
  skip \(\Rightarrow\) (prec (to\u{}a l) a\u{}skip, Some l)
\(|\) sequence i\(\sb{1}\) i\(\sb{2}\) \(\Rightarrow\)
  let (i'1, l') := abstract\u{}i i\(\sb{1}\) l in
  match l' with
    None \(\Rightarrow\) (a\u{}sequence i'1 (prec false\u{}assert (mark i\(\sb{2}\))), None)
  \(|\) Some l' \(\Rightarrow\) let (i'2, l'') := abstract\u{}i i\(\sb{2}\) l' in (a\u{}sequence i'1 i'2, l'')
  end
\(|\) assign x e \(\Rightarrow\)
  (prec (to\u{}a l) (a\u{}assign x e), Some (ab\u{}update l x (ab\u{}eval (ab\u{}lookup l) e)))
\(|\) while b i \(\Rightarrow\)
  match intersect\u{}env true l b with
    None \(\Rightarrow\) 
   (prec (to\u{}a e)(a\u{}while b (a\u{}conj (a\u{}not (a\u{}b b)) (to\u{}a l)) (mark i)), Some l)
  \(|\) Some l' \(\Rightarrow\)
    let (i',l'') := fp l b i (abstract\u{}i i) in
      match l'' with
        None \(\Rightarrow\) (prec (to\u{}a l) (a\u{}while b (to\u{}a l) i'), intersect\u{}env false l)
      \(|\) Some l'' \(\Rightarrow\) (prec (to\u{}a l) (a\u{}while b (to\u{}a l'') i'), intersect\u{}env false l'' b)
      end
  end
end.
\end{alltt}
This abstract interpreter is a programming tool: it can be run with an
instruction and a set of initial approximations for variables. It
returns the same instruction, where each location is annotated with
properties about the variables at this location, together with
properties for the variables at the end.  
This abstract interpreter is structurally recursive and can be run
inside the Coq proof system.

We proved a correctness statement for this abstract
interpreter.  This 
statement relies on the verification condition generator that we
described earlier.
\begin{alltt}\sf
Theorem abstract\u{}i\u{}sound:
  \(\forall\) i e i' e' g, abstract\u{}i i e = (i', e') \(\rightarrow\) i\u{}lc m g (vcg i' (to\u{}a' e'))).
\end{alltt}
This theorem is proved by induction on {\sf i}.  We need to
establish a few facts:
\begin{enumerate}
\item the order of variables does not change in successive abstract
environments,
\item abstract execution is actually monotonic: given wider approximations,
execution yields wider results (given reasonable assumptions for 
{\sf intersect\u{}env})
\item the {\sf fp} function (which handles loop bodies) either yields an
abstract environment that is an over approximation of its input or detects
non-termination of the loop body,
\item the verification condition generator is monotonic with respect
  to implication: if the conditions generated for \(i\) and
  a post-condition \(p\)
  hold and \(p\rightarrow q\) is valid, then the conditions generated
  for \(i\) and \(q\) also hold  and
  {\sf pc \(i\) \(p\) \(\rightarrow\) pc \(i\) \(q\)} is also valid.
  This property is needed because abstract
  interpreters and condition generators work in reverse directions.
\end{enumerate}
This abstract interpreter was developed in a modular fashion, where the
domain of abstract values is described using a module interface.  We
implemented an instance of this domain for intervals.
\section{Conclusion}
This overview of formalized programming language semantics
 is elementary in its choice of a very limited programming language.
Because of this, some important aspects of programming languages
are overlooked: {\em binding}, which appears as soon as local variables or
procedures and functions are allowed, {\em typing}, which is a useful
programming concept for the early detection of programming errors,
{\em concurrency}, which is useful to exploit modern computing
architectures, etc.  Even for this simplistic
programming language, we could also have covered two more aspects:
program transformations
\cite{reasoningexecutable} and compilation \cite{bertot98}.

Three aspects of this work are original: we obtain tools that can be
executed inside the Coq prover for proof by reflection; our work on
denotational semantics shows that the conventional extraction facility
of the Coq system can also be used for potentially non terminating functions,
thanks to well chosen extraction for Tarski's fixpoint theorem;
last, our description of an abstract interpreter is the first to rely on
axiomatic semantics to prove the correctness of an abstract interpreter.

Concerning reflection, we find it exciting that the theorem prover can be
used to execute programs in the object language (in work not reported here
we show how to construct an incomplete interpreter from a structural operational
semantics), to generate
condition verifications about programs (thanks to the verification condition
generator), and to prove the conditions using the normal mode of operation
of theorem prover.  More interestingly, the abstract interpreter can be
run on programs to generate simultaneously annotated programs and the proof
that these annotated programs are consistent.

Formal verification techniques based on verification condition
generators suffer from the burden of explicitely writing the loop
invariants.  Chaieb already suggested that the loop invariants could
be obtained through abstract interpretation \cite{DBLP:conf/ictac/Chaieb06},
generating proof traces that can be verified
in theorem provers.  Our partial correctness theorem for the abstract
interpreter suggests a similar approach here, except that we also
proved the abstract interpreter correct.  An interesting improvement
would be to make manually written assertions collaborate with
automatically generated ones.  First there should be a way to assume that
all assertions computed by an abstract interpreter are implicitly present
in assertions; second abstract interpreters could take manual annotations
as clues to improve the quality of the abstract environments they compute.
\bibliographystyle{plain}
\bibliography{a}

\newcommand{\noopsort}[1]{}
\begin{thebibliography}{10}

\bibitem{g+01:esop}
G.~Barthe, G.~Dufay, L.~Jakubiec, S.~Melo de~Sousa, and B.~Serpette.
\newblock {A Formal Executable Semantics of the JavaCard Platform}.
\newblock In D.~Sands, editor, {\em Proceedings of ESOP'01}, volume 2028 of
  {\em LNCS}, pages 302--319. Springer-Verlag, 2001.

\bibitem{BartheRuysBarendregt95}
Gilles Barthe, Mark Ruys, and Henk Barendregt.
\newblock A two-level approach towards lean proof-checking.
\newblock In {\em TYPES '95: Selected papers from the International Workshop on
  Types for Proofs and Programs}, pages 16--35, London, UK, 1996.
  Springer-Verlag.

\bibitem{DBLP:conf/types/BerghoferN00}
Stefan Berghofer and Tobias Nipkow.
\newblock Executing higher order logic.
\newblock In Paul Callaghan, Zhaohui Luo, James McKinna, and Robert Pollack,
  editors, {\em TYPES}, volume 2277 of {\em Lecture Notes in Computer Science},
  pages 24--40. Springer, 2000.

\bibitem{bertot98}
Yves Bertot.
\newblock A certified compiler for an imperative language.
\newblock Research Report RR-3488, INRIA, 1998.

\bibitem{Bertot00a}
Yves Bertot.
\newblock Formalizing a jvml verifier for initialization in a theorem prover.
\newblock In {\em Computer Aided Verification (CAV'2001)}, volume 2102 of {\em
  LNCS}, pages 14--24. Springer-Verlag, 2001.

\bibitem{Coqart}
Yves Bertot and Pierre Cast{\'e}ran.
\newblock {\em Interactive Theorem Proving and Program Development, Coq'Art:the
  Calculus of Inductive Constructions}.
\newblock Springer-Verlag, 2004.

\bibitem{reasoningexecutable}
Yves Bertot and Ranan Fraer.
\newblock {R}easoning with {E}xecutable {S}pecifications.
\newblock In {\em TAPSOFT'95}, volume 915 of {\em LNCS}, pages 531--545, 1995.

\bibitem{CentaurShort}
Patrick Borras, Dominique {Cl{\'e}ment}, Thierry Despeyroux, Janet Incerpi,
  Gilles Kahn, Bernard Lang, and {Val{\'e}rie} Pascual.
\newblock Centaur: the system.
\newblock In {\em Third Symposium on {S}oftware {D}evelopment {E}nvironments},
  1988.

\bibitem{Boutin97b}
S.~Boutin.
\newblock Using reflection to build efficient and certified decision
  procedures.
\newblock In Martin Abadi and Takahashi Ito, editors, {\em TACS'97}, volume
  1281. LNCS, Springer-Verlag, 1997.

\bibitem{DBLP:conf/ictac/Chaieb06}
Amine Chaieb.
\newblock Proof-producing program analysis.
\newblock In Kamel Barkaoui, Ana Cavalcanti, and Antonio Cerone, editors, {\em
  ICTAC}, volume 4281 of {\em Lecture Notes in Computer Science}, pages
  287--301. Springer, 2006.

\bibitem{NUPRL}
Robert Constable, S.~F. Allen, H.~M. Bromley, W.~R. Cleaveland, J.~F. Cremer,
  R.~W. Harber, D.~J. Howe, T.~B. Knoblock, N.~P. Mendler, P.~Panangaden, J.~T.
  Sasaki, and S.~F. Smith.
\newblock {\em Implementing mathematics with the {Nuprl} proof development
  system}.
\newblock Prentice-{H}all, 1986.

\bibitem{Agda}
Catarina Coquand.
\newblock Agda.
\newblock www.cs.chalmers.se/\~{\null}catarina/agda.

\bibitem{CousotCousot77-1}
P{.} Cousot and R{.} Cousot.
\newblock Abstract interpretation: a unified lattice model for static analysis
  of programs by construction or approximation of fixpoints.
\newblock In {\em Conference Record of the Fourth Annual ACM SIGPLAN-SIGACT
  Symposium on Principles of Programming Languages}, pages 238--252, Los
  Angeles, California, 1977. ACM Press, New York, NY.

\bibitem{PichardieThesis}
Pichardie David.
\newblock {\em Interpr{\'e}tation abstraite en logique intuitionniste~:
  extraction d'analyseurs Java certifi{\'e}s}.
\newblock PhD thesis, Université Rennes~1, 2005.
\newblock In french.

\bibitem{Dijkstra76}
Edsger~W. Dijkstra.
\newblock {\em A discipline of Programming}.
\newblock Prentice Hall, 1976.

\bibitem{Coq}
Gilles Dowek, Amy Felty, Hugo Herbelin, {G\'erard} Huet, Chet Murthy, Catherine
  Parent, Christine Paulin-Mohring, and Benjamin Werner.
\newblock {\em The Coq Proof Assistant User's Guide}.
\newblock INRIA, May 1993.
\newblock Version 5.8.

\bibitem{filliatre98}
Jean-Christophe Filli{\^a}tre.
\newblock Proof of imperative programs in type theory.
\newblock In {\em International Workshop {TYPES'98}}, volume 1657 of {\em
  Lecture Notes in Computer Science}. Pringer-Verlag, March 1998.

\bibitem{HOL}
Michael J.~C. Gordon and Thomas~F. Melham.
\newblock {\em Introduction to {HOL} : a theorem proving environment for
  higher-order logic}.
\newblock Cambridge University Press, 1993.

\bibitem{DBLP:books/sp/Gordon79}
Michael J.~C. Gordon, Robin Milner, and Christopher~P. Wadsworth.
\newblock {\em Edinburgh LCF}, volume~78 of {\em Lecture Notes in Computer
  Science}.
\newblock Springer, 1979.

\bibitem{hoare69}
Charles Anthony~Richard Hoare.
\newblock An axiomatic basis for computer programming.
\newblock {\em Communications of the ACM}, October 1969.

\bibitem{natural}
Gilles Kahn.
\newblock {N}atural {S}emantics.
\newblock In K.~Fuchi and Maurice Nivat, editors, {\em {P}rogramming of
  {F}uture {G}eneration {C}omputers}. North-Holland, 1988.
\newblock (also appears as INRIA Report no. 601).

\bibitem{Letouzey2002types}
Pierre Letouzey.
\newblock A new extraction for {Coq}.
\newblock In Herman Geuvers and Freek Wiedijk, editors, {\em TYPES 2002},
  volume 2646 of {\em LNCS}. Springer-Verlag, 2003.

\bibitem{MuellerNvOS99}
Olaf M\"uller, Tobias Nipkow, David~von Oheimb, and Oskar Slotosch.
\newblock {HOLCF = HOL + LCF}.
\newblock {\em Journal of Functional Programming}, 9:191--223, 1999.

\bibitem{DBLP:journals/fac/Nipkow98}
Tobias Nipkow.
\newblock Winskel is (almost) right: Towards a mechanized semantics.
\newblock {\em Formal Asp. Comput.}, 10(2):171--186, 1998.

\bibitem{Paulson:Isabelle}
Lawrence~C. Paulson and Tobias Nipkow.
\newblock {\em Isabelle : a generic theorem prover}, volume 828 of {\em Lecture
  Notes in Computer Science}.
\newblock Springer-Verlag, 1994.

\bibitem{Ter95short}
D.~Terrasse.
\newblock Encoding natural semantics in {Coq}.
\newblock In {\em AMAST'95}, Springer-Verlag {LNCS}, July 1995.

\bibitem{LOOPCompiler}
Joachim van~den Berg and Bart Jacobs.
\newblock The loop compiler for java and jml.
\newblock In {\em TACAS 2001}, pages 299--312. Springer-Verlag, 2001.

\bibitem{Oheimb2000}
David von Oheimb.
\newblock {\em Analyzing Java in {Isabelle/HOL}, Formalization, Type Safety,
  and Hoare Logic}.
\newblock PhD thesis, Technische Universit{\"a}t M{\"u}nchen, 2000.

\bibitem{Winskel93}
Glynn Winskel.
\newblock {\em The Formal Semantics of Programming Languages, an introduction}.
\newblock Foundations of Computing. The MIT Press, 1993.

\end{thebibliography}
\end{document}